\documentclass[11pt]{article}
\usepackage{moriond,epsfig}
\usepackage{amsmath}
\usepackage{amssymb}
\usepackage{epsfig}
\usepackage{wrapfig}
\usepackage{array}

\def\be{\begin{equation}}
\def\ee{\end{equation}}
\def\bea{\begin{eqnarray}}
\def\eea{\end{eqnarray}}

\begin{document}
\vspace*{4cm}
\title{PREDICTIONS OF PHYSICAL OBSERVABLES FROM MINIMAL NEUTRINO STRUCTURES}

\author{T. HAMBYE}

\address{Scuola Normale Superiore and INFN Sezione di Pisa \\ 
Piazza dei Cavalieri 7, I--56126 Pisa, Italy}

\maketitle\abstracts{We find all possible seesaw textures which can
  describe in a natural way the currently observed neutrino oscillation 
  pattern in terms 
  of a minimum number of parameters.
  Natural here means due only to the relative smallness (vanishing) of
  some parameters in the relevant lagrangian, without special
  relations or accidental cancellations among them.
  The corresponding predictions for the mixing angle $\theta_{13}$
  and the effective mass $m_{ee}$ are given.}

\section{Introduction}

In the last five years the neutrino physics has seen remarkable 
experimental developments. Observations 
of atmospheric \cite{sk}, solar \cite{solar},
beam \cite{k2k} and 
reactor \cite{kamland:02a} neutrino deficits have established that 
at least two of the neutrinos have a mass. All these results can 
be nicely accommodated in the simple 3 
neutrino ($\nu_e, \nu_\mu, \nu_\tau$) 
framework with 2 mass differences and 3 mixing angles, determined 
with the values: 
\begin{equation}
  \label{eq:obs}
  \begin{split}
2.3 \cdot 10^{-3}\,\hbox{eV}^2 < 
\delta m^2_{32} & <  3.1 \cdot 10^{-3} \,\hbox{eV}^2\,, 
\quad 6.4 \cdot 10^{-5} \, \hbox{eV}^2 < \delta m^2_{21} < 
7.8 \cdot 10^{-5} \,\hbox{eV}^2 \,,\\ 
0.7 <  \tan^2\theta_{23} & <
  1.3\,, \quad \,\,
  0.35 < \tan^2 \theta_{12} < 0.55\,, \quad \,\,
  \sin^2 \theta_{13}  \lesssim 0.06\,. 
\end{split} 
\end{equation}
A real pattern of neutrino masses and mixings begins to be therefore at hand.
Moreover, in the future, 3 additional parameters
could be measured if not too tiny:
the mixing angle $\theta_{13}$
in neutrino
superbeam and factory experiments \cite{Apollonio:02a},
the CKM type CP-violating phase $\delta$ 
at neutrino factories and the $m_{ee}$ 
combination of masses and 
mixings in the $0\nu2\beta$ decay experiments.
For a hierarchical spectrum of neutrino masses, either
``normal'' ($m_3\gg m_2> m_1$) or ``inverted'' ($m_1\simeq m_2\gg
m_3$), as we shall consider in the following \footnote{The degenerate
  spectrum is not considered since it does not satisfy the naturalness
  criterion as defined below.}, one can tentatively assume that all 
the six oscillation observables
will be measured, some of them with significant precision, perhaps
together with the $0\nu2\beta$ mass $m_{ee}$. This makes a total of seven
observables, which a theory of neutrino masses should be able to
predict. In view of future experiments it would be
crucial that, at the least, we could predict $\theta_{13}$, $m_{ee}$ 
or $\delta$ from the values of the other parameters.

Nevertheless, to predict such experimentally testable 
relations between these seven 
observables turns out to be a very difficult task.
The seesaw mechanism provides a natural explanation for the smallness
of the neutrino masses but it does not explain the above pattern
of neutrino masses and mixings.
A simple way to illustrate this problem is to perform a 
counting of the parameters in the seesaw extended standard model 
based on the following lagrangian with three heavy right handed Majorana 
neutrinos ${N_i}_R$ (after electroweak symmetry breaking):
\begin{equation}
{\cal L} \owns -\frac{1}{2} N^{T}_{R} M_R N_{R} 
- v \bar{N}_{R} Y_\nu N_{L}  + h.c.\,,
\label{lagr}
\end{equation}
where $Y_\nu$
is the neutrino Yukawa coupling matrix, with $N_R^T=(N_1,N_2,N_3)_R$, 
$N_L^T=(\nu_e, \nu_\mu, \nu_\tau)_L$, and $v=174$~GeV. The associated
neutrino mass matrix in the flavour basis is given by
\begin{equation}
M_\nu= - U_l^T Y^T_\nu M_R^{-1} Y_\nu U_l v^2 \,,
\label{Mnu}
\end{equation}
with $U_l$ the mixing matrix resulting from the diagonalization
of the charged lepton mass matrix.
In the basis where the charged lepton and right handed neutrino mass matrices
are real and diagonal, in addition to the three right handed neutrino 
masses, there are 9 real parameters and 6 phases in the 
Yukawa coupling matrix $Y_\nu$, which 
give a total of 18 unknown parameters. These 18 parameters 
must necessarily be known in order
to really test the seesaw mechanism
and the underlying flavour structure, and there are only seven observables!

A simple minded conjecture which allows to correlate some of the 
7 observables, is that these correlations could arise from the economy
in the number of independent basic parameters.
As the simplest example, here one can consider 
the possibility that the lagrangian 
has a maximum
number of negligible small entries; we insist on the fact that
the vanishing parameters have to be those of the basic lagrangian, not 
of the neutrino mass matrix which in the seesaw model are only combinations 
of the fundamental parameters.
Such a conjecture
may offer an opportunity to solve
the above flavour problem. We believe in particular that to address
this difficult problem, it is important 
to determine in a systematic way how complex
must be the minimal flavour structures to explain the known data.
More generally this also provides one of the rare 
chance to test indirectly 
the seesaw origin of the neutrino masses, because in this case
the correlations obtained will be in general closely related to 
the seesaw structure of the neutrino mass matrix. Note that there are
a priori several mechanisms, including symmetries, which 
could be responsible for the 
relative smallness (or vanishing) of some parameters. The
purpose of this work is not to identify these mechanisms but 
to see what are the consequences of assuming that such a mechanism exists.

To find all correlations between observables assuming vanishing entries in 
the basic seesaw lagrangian would require examining a
huge number of different possibilities. In the following we
shall consider only the possibilities describing the currently
observed pattern of the data~(\ref{eq:obs}) in a ``natural'' way,
i.e.~only by the relative smallness (vanishing) of some parameters in
${\cal L}$, barring special relations or accidental cancellations among
them. In particular this should be the case when accounting
simultaneously for the smallness 
of $R\equiv \delta m^2_{21}/\delta m^2_{32}$ and for
the largeness of $\theta_{23}$, the most peculiar feature of the data so
far, even though an accidental cancellation might also produce the
same feature.

\section{The number of parameters and $U_l$ rotation issues}

Assuming vanishing 
entries in Eq.~\ref{lagr}, one important preliminary remark which must 
be done is that there are always more parameters in
the basic seesaw lagrangian than in the light neutrino mass matrix.
This comes from the fact that $M_\nu$ depends only on $\sim Y_\nu^2/M_N$ 
and not on the normalization of the lines of $Y_\nu$ and of the $M_N$
separately. Formally writing $Y_\nu$ and $M_N$ as
\begin{equation}
Y_\nu=D A \,, \,\,\,\,\,\,\,\, M_N= D \mu D \,,
\label{Amu}
\end{equation} 
with $D=\hbox{diag}(d_1,d_2,d_3)$ an adimensional diagonal 
matrix taken in such a way that the lines of $A$ are
normalized to one, this manifests itself by the 
fact that, for fixed values of $A$ and $\mu$, the neutrino mass 
matrix is independent of $D$.
And what determines the number of correlations among physical 
observables is the number
of independent ``effective'' parameters 
in the neutrino mass matrix, not in the lagrangian. Since there 
are 6 real observables, if
the neutrino mass matrix has $n$ effective real parameters, there 
will be $6-n$ relations between the parameters.
In practice one can convince oneself that,  
in order to reproduce the known data on $\delta m^2_{23}$, 
$\delta m^2_{12}$, $\theta_{23}$, $\theta_{12}$ and $\theta_{13}$ along the 
assumptions we made, we 
need at least 4 effective parameters. In the following
we have determined all minimal (i.e.~with 4 
effective parameters) configurations which are not already excluded by
experiment, that is to say which give 2 predictions, one for $\theta_{13}$ and 
one for $m_{ee}$ as a function of $\delta m^2_{23}$, 
$\delta m^2_{12}$, $\theta_{23}$ and $\theta_{12}$ (and in one case also 
as a function of the phase $\delta$). 

\begin{table}[b!]
  \centering
\caption{Summary of the possible correlations between $\theta_{13}$,
  $m_{ee}$ and $\theta_{23}$, $\theta_{12}$, $\delta$,
  $R=\delta m^2_{32}/ \delta m^2_{21}$
  (with $m_{\text{atm}}\equiv\sqrt{|\delta m^2_{32}|}$). The 
  columns ``$N_R$'' and ``$U_l$'' give 
  the number of right handed neutrinos involved in the see-saw
  realization of each case and the form of the $U_l$ rotation matrix. An 
  inverse hierarchy is obtained only in
  the case E. Cases A2, B2, E2, are obtained from A1, B1, E1, with 
  the replacements $\tan\theta_{23} \rightarrow \cot\theta_{23}$ and 
  $\cos \delta \rightarrow - \cos\delta$.} 
\begin{equation}
 \begin{array}{|c||c|c|c|c|c|} 
     \hline
    &
    \sin\theta_{13} &
    |m_{ee}|/m_{\text{atm}} &
    N_R &
    U_l \\
    \hline\hline
    \text{A1}&
    \displaystyle\rule[-0.4cm]{0cm}{1.1cm}
    \frac{1}{2}\tan\theta_{23}\sin2\theta_{12}\,\sqrt{R} &
    \sin^2\theta_{12}\sqrt{R} &
    \text{2} &
    \mathbf{1} \\
    \hline
    \text{B1} &
    \displaystyle\rule[-0.4cm]{0cm}{1.1cm}
    \frac{1}{2}\tan\theta_{23}\tan2\theta_{12}\,(R\cos2\theta_{12})^{1/2} &
    0 &
    \text{3} &
    \mathbf{1} \\
    \hline
    \text{C} &
    \displaystyle\rule[-0.4cm]{0cm}{1.1cm}
    \frac{1}{2}\tan2\theta_{12}\,(R\cos2\theta_{12})^{3/4} &
    0 &
    \text{3} &
    R_{23} \\
    \hline
    \text{D} &
    \displaystyle\rule[-0.4cm]{0cm}{1.1cm}
    \frac{1}{2}\frac{\tan2\theta_{12}}{|\tan2
    \theta_{23}|}(R\cos2\theta_{12})^{1/2} & 
    \displaystyle\rule[-0.4cm]{0cm}{1.1cm}
    \left(\frac{\sin\theta_{13}}{\cos2\theta_{23}}\right)^2 &
    \text{3} &
    \mathbf{1} \\
    \hline
    \hline
    \text{E1} &
    \displaystyle\rule[-0.4cm]{0cm}{1.1cm}
    -\frac{\tan\theta_{23}}{\cos\delta}\frac{1-\tan\theta_{12}}{1+
    \tan\theta_{12}} &
    2\cot\theta_{23}\sin\theta_{13} &
    \text{2, 3} &
    R_{12}(R_{23}) \\
    \hline
 \end{array}
\nonumber
\end{equation}
  \label{tab:results}
\end{table}

Before presenting the results, note also that if we restrict 
ourselves to 4 effective parameters, the charged lepton mixing 
matrix must have a simple structure. The only forms which are 
possible \cite{bhr} are a simple rotation along one of the 
flavour $e$, $\mu$ or $\tau$ axis, i.e.
$U_l=R_{23}$, $R_{12}$, $R_{13}$,
or a double rotation $U_l=R_{12} R_{23}$, $R_{13} R_{23}$. Any 
other rotation would bring too many effective parameters to the 
neutrino mass matrix or, like for example $R_{23} 
R_{12}$, would lead to an already excluded value of $\theta_{13}$.

\begin{table}[t]
  \centering
\caption{Parameters for the see-saw cases with 2 right handed
  neutrinos of Table~\ref{tab:results}.  $A$ is the Dirac neutrino mass 
matrix with $(A A^\dagger)_{ii}$ normalized to unity and $\mu$ is related 
to the right handed 
neutrino mass matrix by Eq.~\ref{Amu}.
The lagrangians leading to the corresponding predictions of Table 1
are given by injecting $A$ and $\mu$ below in Eqs.~\ref{lagr} 
and \ref{Amu} 
with $D$ any diagonal matrix and with $\mu_0$ any mass scale. 
$\epsilon$ 
and $\sigma$ denote
small entries relative to unity. c, s or c', s' denote the cosine and the
sine of arbitrary angles, $\theta$ and $\theta'$. $U_l$ is 
the rotation matrix of the left handed charged leptons. Case E1 (E2) can
also be obtained from the E1 (E2) configurations below with $c=1$, $s=0$,
$U_l=R_{12} R_{23}$ ($U_l= R_{13} R_{23}$) and with the small entry in
$\mu_{11}$, $\mu_{22}$, $A_{12}$ or $A_{21}$.}
\begin{equation} 
\begin{array}{|c||c|c|c|} 
   \hline
    &
    \mu/\mu_0 &
    A &
    U_l \\
    \hline\hline
    \text{A1} & 
    \begin{pmatrix} \epsilon & 0 \\ 0 & 1 \end{pmatrix} &
    \begin{pmatrix} 0 & s & c \\ c' & s'e^{i\phi} & 0 \end{pmatrix} &
    \mathbf{1} \\
    \text{A2} &
    \begin{pmatrix} \epsilon & 0 \\ 0 & 1 \end{pmatrix} &
    \begin{pmatrix} 0 & s & c \\ c' & 0 & s'e^{i\phi} \end{pmatrix} &
    \mathbf{1} \\
    \hline
    \text{E1 (E2)} &
    \begin{pmatrix} 0 & 1 \\ 1 & 0 \end{pmatrix} &
    \begin{pmatrix} 1 & 0 & 0 \\ 0 & c & s \end{pmatrix} &
    \text{$R_{12}$ ($R_{13}$)} \\
    & \multicolumn{2}{c|}{\text{+ 1 small entry}} & \\
    \hline
\end{array}
\nonumber
\end{equation}
 \label{tab:SSC2}
\end{table}

\section{Results}

Following a methodology described in Ref.~\cite{bhr}, we find that there are
only five definite testable sets of predictions not already excluded by 
the data, given in Table 1.
There are
only four cases (A, B, C, D) that allow to connect $\sin\theta_{13}$ and
$m_{ee}$ with $\theta_{23}$, $\theta_{12}$ and 
$R \equiv \delta m^2_{32}/ \delta m^2_{21}$, whereas in case E the
correlation also involves the CP-violating phase $\delta$.
The relations quoted in Table~\ref{tab:results} are obtained through
an expansion in $R$. The higher orders in the expansion are suppressed
by $R^{1/2}$ in all cases except $E$, where the leading corrections to
the relations in Table~\ref{tab:results} are of order $R$.  In all
cases, anyhow, the exact relations can be obtained from
Tables~\ref{tab:SSC2} and \ref{tab:SSC3}, where the
sets of parameters that originate these correlations are shown. 
Table~\ref{tab:results}
specifies the number of right handed neutrinos involved and it also gives
the form of the rotation on the charged lepton sector.
For the cases with two neutrinos, the third neutrino is assumed
to be very massive and/or to have negligible couplings.
E is the only case that leads to an ``inverted''
spectrum. For cases A1, B1, E1, the independent possibility exists
where $\tan\theta_{23} \rightarrow \cot\theta_{23}$, $\cos\delta \rightarrow
-\cos\delta$, denoted in the following by A2, B2, E2, respectively.
Case A is discussed in
Ref.~\cite{Frampton:02b,Raidal:02a}.The prediction for $\theta_{13}$ in 
case B was already
obtained approximately from other 
phenomenological assumptions \cite{ak}.
For related works see also Refs.\cite{div} and references therein.

\begin{figure}[t]
\begin{center}
\epsfig{file=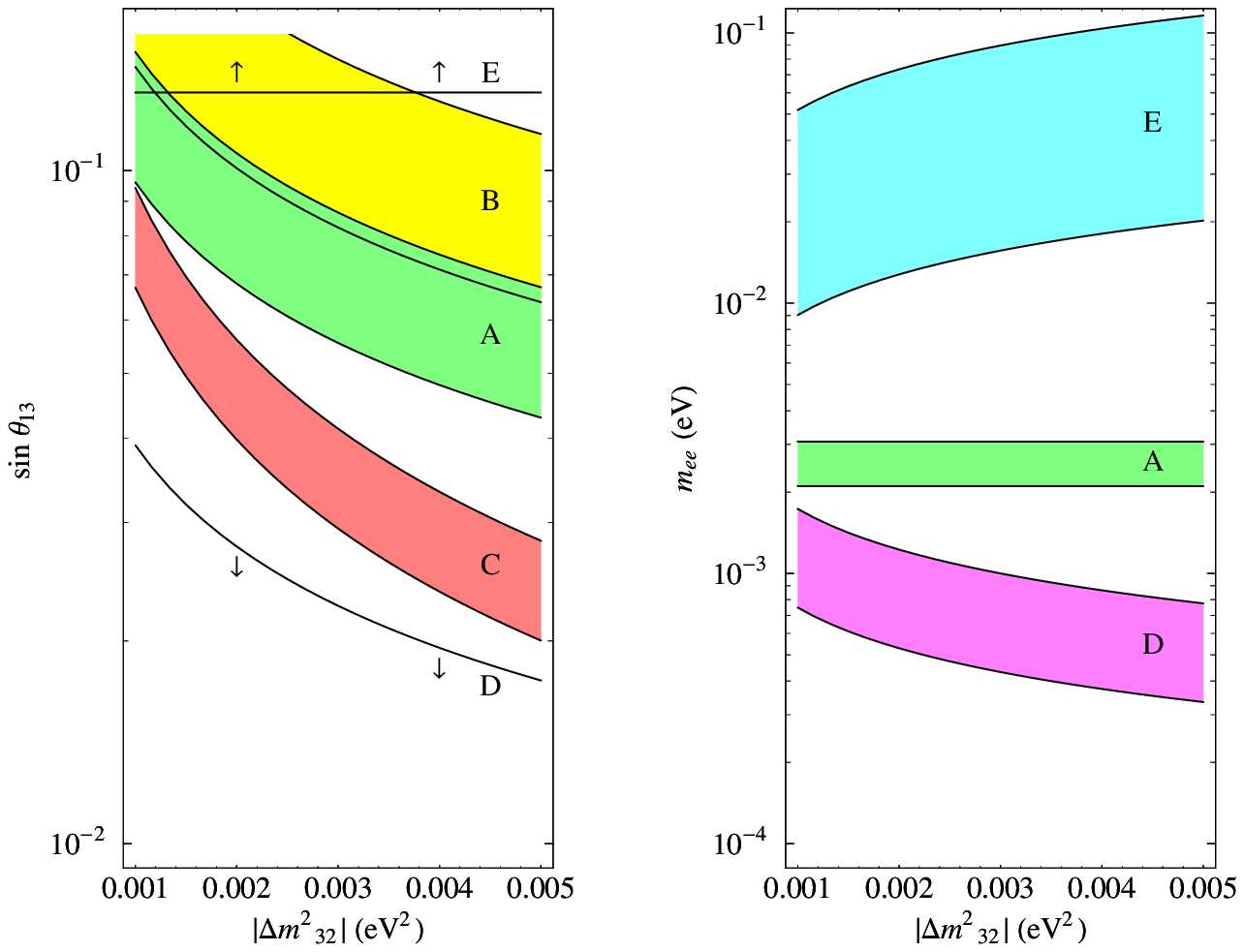,width=0.84\textwidth}
\end{center}
\hspace*{0.05\textwidth}
\begin{minipage}[t]{0.445\textwidth}
  \caption{Ranges of values for $\sin\theta_{13}$ at 90\% confidence
    level for the different cases, plotted as a function 
    of $\delta m^2_{32}$.
    Cases D, E, which only give a bound on $\sin\theta_{13}$, are shown 
    with a double arrow. \label{fig:s13}}
\end{minipage}\hspace*{0.03\textwidth}
\begin{minipage}[t]{0.440\textwidth}
  \caption{Ranges of values for $m_{ee}$ at 90\% confidence
  level. Cases B and C predict $m_{ee}=0$.
\label{fig:mee}} 
\end{minipage}
\end{figure}

Given the present knowledge of $\theta_{23}$, $\theta_{12}$ 
and $\delta m^2_{21}$,
including the recent Kamland result~\cite{kamland:02a}, the ranges of
values for $\sin\theta_{13}$ are shown in Fig.~\ref{fig:s13} at 90\%
confidence level for the different cases.  It is interesting that all
the ranges for $\sin\theta_{13}$, except in case D, are above $\simeq0.02$
and some can saturate the present limit.  Long-baseline experiments 
should explore a significant portion of
this range while reducing at the same time the uncertainties of the
different predictions at about 10\% level~\cite{Apollonio:02a}.  Note
that, in cases E, although the determination of $\sin\theta_{13}$ requires
the knowledge of the CP violating phase as well, the allowed range is
still limited, being $\sin\theta_{13}\gtrsim 0.10$.  Furthermore, the
requirement of not exceeding the present experimental bound on
$\sin\theta_{13}$ gives a lower bound on $|\cos\delta|$ (and therefore an
upper limit on CP-violation) that we can quantify as
$|\cos\delta| > 0.8$ (at 90 \% CL)
given the present uncertainties. Notice that $\cos\delta < 0$ ($>0$)
in case E1 (E2). Verifying the prediction for $\sin\theta_{13}$ in case D
would require the measurement of $\theta_{23} \neq 45^\circ$; a bound on
$|1-\sin^22\theta_{23}|$ only sets an upper bound 
on $\sin\theta_{13}$, as shown
in Fig.~1.

While the predictions for $\sin\theta_{13}$ are in an experimentally
interesting range, the expectations for the $0\nu2\beta$-decay
effective mass are mostly on the low side, except, as
expected~\cite{Feruglio:02a}, in the only inverted hierarchical case
E. The ranges for each individual cases with non-vanishing $m_{ee}$
are shown in Fig.~\ref{fig:mee}.  The challenge of detecting a non
zero $m_{ee}$, when applicable, is therefore harder than for
$\sin\theta_{13}$, with a better chance for the only inverted hierarchical
case E.

Note that we also found 3 configurations which predict 
$\theta_{13}=m_{ee}=0$ which we don't show because they are 
not testable. Therefore, if for a large majority of the configurations
an experimentally
accessible value of $\sin \theta_{13}$ is found, the existence of 
these configurations does not allow us to say 
that such a large value is a prediction of our assumptions.

Finally note that in a model like the triplet seesaw model \cite{hs}, 
the neutrino 
mass matrix is a fundamental quantity since it is directly proportional
to the lepton-lepton-triplet coupling matrix.
In this case, following our assumptions, we find \cite{bhr} that there are 
only two sets of relations, those of cases C and E, which 
are testable and not excluded by the present data.

\section{Summary}

\begin{table}[t]
  \centering
\caption{Parameters for the see-saw cases with 3 right handed
  neutrinos of Table~\ref{tab:results}. $\epsilon$ and 
$\sigma$ denote small positive quantities, while $a={\cal O}(1)$. $\mu$,
$A$ and $U_l$ as in Table~\ref{tab:SSC2}. Case E1 (E2) can
also be obtained from the E1 (E2) configurations below with $c=1$, $s=0$
and $U_l=R_{12} R_{23}$ ($U_l= R_{13} R_{23}$).}
\begin{equation}
{\small
\begin{array}{|c||c|c|c|} 
    \hline
    &
    \mu/\mu_0 &
    A &
    U_l \\
    \hline\hline
    \text{B1 (B2)} &
    \begin{pmatrix}
      a & c & s \\ c & 0\,(\epsilon e^{i\phi}) & 0 \\ 
      s & 0 & \epsilon e^{i\phi}\, (0)
    \end{pmatrix} &
    \begin{pmatrix}
      1 & 0 & 0 \\ 0 & 1 & 0 \\ 0 & 0 & 1
    \end{pmatrix} &
    \mathbf{1} \\
    \text{} &
    \begin{pmatrix}
      0 & c & s \\ c & 0\,(\sigma) & 0 \\ s & 0 & \sigma \,(0)
    \end{pmatrix} &
    \begin{pmatrix}
      c' & s'e^{i\phi} & 0 \\ 0 & 1 & 0 \\ 0 & 0 & 1
    \end{pmatrix} \!,\;
    \begin{pmatrix}
      c' & 0 & s'e^{i\phi} \\ 0 & 1 & 0 \\ 0 & 0 & 1
    \end{pmatrix} &
    \mathbf{1} \\
    \text{} &
    \begin{pmatrix}
      ae^{i\phi} & 1 & 0 \\ 1 & 0 & 0 \\ 0 & 0 & \sigma
    \end{pmatrix} &
    \begin{pmatrix}
      1 & 0 & 0 \\ 0 & 1 \, (0) & 0 \, (1) \\ 0 & s & c
    \end{pmatrix} &
    \mathbf{1} \\
    \text{} &
    \begin{pmatrix}
      0 & 1 & 0 \\ 1 & 0 & 0 \\ 0 & 0 & \sigma
    \end{pmatrix} &
    \begin{pmatrix}
      c' & s'e^{i\phi} & 0 \\ 0 & 1\,(0) & 0\, (1) \\ 0 & s & c
    \end{pmatrix} \!,\;
    \begin{pmatrix}
      c' & 0 & s'e^{i\phi} \\ 0 & 1 \, (0) & 0 \, (1) \\ 0 & s & c
    \end{pmatrix} &
    \mathbf{1} \\
    \hline
    \text{C} &
    \begin{pmatrix}
      0 & 1 & 0 \\ 1 & 0 & 0 \\ 0 & 0 & \sigma
    \end{pmatrix} &
    \begin{pmatrix}
      c' & 0 & s'e^{i\phi} \\ 0 & 1 & 0 \\ 0 & 0 & 1
    \end{pmatrix} &
    R_{23} \\
    \hline
     \text{D} &
    \begin{pmatrix}
      a & c & s \\ c & 0 & \epsilon e^{i\phi} \\ s & \epsilon
      e^{i\phi} & 0
    \end{pmatrix} &
    \begin{pmatrix}
      1 & 0 & 0 \\ 0 & 1 & 0 \\ 0 & 0 & 1
    \end{pmatrix} &
    \mathbf{1} \\
    \text{} &
    \begin{pmatrix}
      0 & c & s \\ c & 0 & \sigma \\ s & \sigma & 0
    \end{pmatrix} &
    \begin{pmatrix}
      c' & s'e^{i\phi} & 0 \\ 0 & 1 & 0 \\ 0 & 0 & 1
    \end{pmatrix} \!,\;
    \begin{pmatrix}
      c' & 0 & s'e^{i\phi} \\ 0 & 1 & 0 \\ 0 & 0 & 1
    \end{pmatrix} &
    \mathbf{1} \\
    \hline
    \text{E1 (E2)} &
    \begin{pmatrix}
      0 & \sigma & 0 \\ \sigma & 0 & 0 \\ 0 & 0 & 1
    \end{pmatrix} &
    \begin{pmatrix}
      0 & c & s \\ 1 & 0 & 0 \\ 1 & 0 & 0
    \end{pmatrix}\!,
    \begin{pmatrix}
      0 & c & s \\ 1 & 0 & 0 \\ 0 & 1 & 0
    \end{pmatrix}\!,
    \begin{pmatrix}
      0 & c & s \\ 1 & 0 & 0 \\ 0 & 0 & 1
    \end{pmatrix} &
    R_{12} \, (R_{13})\\
    \hline
\end{array}
}
\nonumber
\end{equation}
\label{tab:SSC3}
\end{table}

The economy in the number of basic parameters could be at the origin
of some correlations between the physical observables in the neutrino
mass matrix. At the present state of knowledge, the variety of the
possibilities for the basic parameters themselves is large.
Finding the minimal cases that describe the present pattern of the
data in a natural way could be a first step in the direction of
discriminating the relevant ${\cal L}$. This we have done with the results
summarized in Tables~\ref{tab:results}--\ref{tab:SSC3} and illustrated
in Figs.~1 and 2. It is remarkable that the number of
possible correlations between the physical observables is limited
(Table~\ref{tab:results}), with a relatively larger number of
possibilities for the basic parameters
(Tables~\ref{tab:SSC2}--\ref{tab:SSC3}). The relatively best chance
for selecting experimentally one out of the few relevant cases is
offered by $\sin\theta_{13}$. Combining this with independent
studies of leptogenesis or of lepton flavour violating effects could
lead to the emergence of an overall coherent picture.
We have insisted on ``naturalness'' both in solving the ``large
$\theta_{23}$-small $R$'' problem for the normal hierarchy case and in
obtaining a significant deviation of $\theta_{12}$ from $45^\circ$, with
small $R$, in the inverted hierarchy case. Explaining these
features in a natural way offers a possible interesting guidance for
model building.

\section*{Acknowledgments}
It is a pleasure to thank R.~Barbieri and A.~Romanino with whom 
this work \cite{bhr} has 
been done. This work was 
supported by the TMR, EC-contract No. HPRN-CT-2000-00152.

\end{document}